# Metal-coated carbon nanotube tips for Magnetic Force Microscopy

Zhifeng Deng[1,4], Erhan Yenilmez[2,4], Josh Leu[1,4], J.E. Hoffman[2,4], Eric Straver[2,4], Hongjie Dai[3,4], Kathryn A. Moler[1,2,4]

1 Department of Physics, Stanford University, Stanford, California 94305

2 Department of Applied Physics, Stanford University, Stanford, California 94305

3 Department of Chemistry, Stanford University, Stanford, California 94305

4 Geballe Laboratory for Advanced Materials, Stanford University, Stanford, California 94305

**Abstract:**

*We fabricated cantilevers for magnetic force microscopy with carbon nanotube tips coated with magnetic material. Images of a custom hard drive demonstrated 20 nm lateral resolution, with prospects for further improvements.*

Nano-magnetic materials[1,2] and the shrinking bit size of magnetic recording media[3] increase demand for Magnetic Force Microscopy (MFM) with high spatial resolution. An MFM probe consists of a magnetic tip mounted on a cantilever. The resolution is limited by the tip's size, shape, and height above the sample surface. Recent innovative MFM probes include focused ion beam (FIB) milled sharp tips, which reportedly resolved sub-30nm wide one dimensional features;[4] tips with an FIB-drilled hole, which imaged 50nm transitions on a hard drive;[5] tips with magnetic dots deposited through stencil masks, which imaged features with 42 nm FWHM;[6] and electron beam deposition (EBD) tips.[7] Recently reported nanotube-based MFM tips used magnetic catalyst particles[8,9] to image 85 nm features. These nanotube tips, however, were typically not suitable for topography because they could not be shortened,[8] and the amount of magnetic material at the tip was not controllable. There is still need for easily fabricated non-invasive MFM probes with high-resolution capability. Here we show the development of metal-coated carbon nanotube (CCNT) probes to address this need (Fig. 1).

Our probes improve MFM imaging in several ways. The magnetic material on the CCNT tip is contained in a narrow cylinder, in contrast to standard pyramidal tips. The smaller volume of magnetic material that contributes to the signal can result in higher spatial resolution. The sharper tip radius of the CCNT also allows a better topographic scan, which yields a more accurate height profile for the tip to follow. CCNT tips are less invasive than commercially available tips because a smaller amount of magnetic material is brought into close proximity with the sample. Finally, the magnetic material is distributed on the tip along an easy-to-model straight cylinder. If the cylinder is sufficiently long, interactions between the geometrically difficult-to-model pyramid tip and the sample are negligible. This feature may result in more quantitative MFM studies.

Our process consists of the growth of carbon nanotubes using wafer-scale chemical vapor deposition on commercial silicon tips for tapping mode atomic force microscopy,[10] followed by nanotube shortening and deposition of metal films. Tapping mode cantilevers are not as damaging to the sample or the tip as contact mode cantilevers, and allow us to finely control the tip-sample separation. A few angstroms of iron are deposited on the cantilever and the pyramidal tip in an e-beam metal evaporator.[11] The tips are placed inside a CVD furnace, heated up to 900°C in Ar and fed a mixture of methane, ethylene and hydrogen gases for 7 min. The iron particles catalyze the self-assembly of carbon nanotubes. We almost always find a few tubes near the apex of the tip. The carbon nanotubes grown by this process have single or double walls, diameters of 1 to 5nm, and a wide distribution of lengths and angles relative to the pyramidal tip. We shorten the long nanotubes protruding from the apexes of the pyramids to a few hundred nm by an electrical cutting method.[10] Then the nanotube tips are coated inside an e-beam evaporator (Fig. 1):[12] first a sticking layer of titanium, followed by a layer of cobalt, and finally a layer of titanium to prevent the oxidation of the magnetic material. Typical resulting tips are shown in Fig. 2. In this work, we found that 3 nm Ti/7 nm Co/3 nm Ti gave good results, but the choice of materials and thicknesses has not been fully optimized.

We have tried CCNT tips with different lengths. In general, shorter nanotubes give better topography scans because they are stiffer.[13] However, when the nanotube is much shorter than ~250nm, we find that the magnetic material on the pyramid of the tip starts to contribute significantly to the signal, decreasing our magnetic spatial resolution. Thus, we prefer to work with nanotubes with lengths around 250nm (Fig. 2).

The spring constant of the chosen cantilevers is about 5N/m; the resonance frequency is around 120kHz; and the quality factor in air is around 100. To test the CCNT cantilevers, we employ the Tapping/Lift™ mode of a Digital Instruments Nanoscope III SPM at room temperature and in air to map the phase shift of the cantilever while scanning a sample. Imaging parameters such as lift height and amplitude of oscillation are optimized to yield the best resolution with each tip. Prior to imaging, we magnetize the tip parallel to the nanotube using a strong permanent magnet. There is no other external magnetic field applied.

The images shown here are images of custom written tracks on a perpendicular recording media (Fig. 3). The linear density of the tracks varies from 100 to 1000 kilo flux changes per inch (kfci), or 254 to 25.4 nm/flux change. Before the tracks are written, the surface is erased, using a DC erase signal on some parts of the disk and an AC signal on others. The images in Fig. 3 are from a DC-erased section. The surface roughness of the sample is 0.3 nm rms. We choose this sample for its combination of flat topography and small magnetic features of a known size.

Fig. 3(a) shows six tracks with decreasing feature size, imaged with the CCNT shown in Fig. 2(a). This tip was one of our earliest efforts. It was metallized with a relatively thick film, and has a diameter of 28 to 30nm, or a radius of curvature of ~15nm. The cantilever had a resonance frequency of 81kHz. It was able to image the fine details of the bits of the densest track, as shown in Fig. 3(b). These images were from a DC-erased section of the hard drive. Comparable images were obtained using a commercial MFM tip.

CCNT tips with smaller diameters yielded better spatial resolution. The tip shown in Fig. 2(b), which was one of our typical best tips, had a diameter of 15 nm as estimated from the SEM picture. Fig. 4(a) shows an MFM image of tracks in a DC-erased area taken with a commercially available tip. Fig. 4(b) shows an image of two tracks in a DC-erased area taken with the CCNT tip from Fig. 2(b). The CCNT tip showed finer details than the commercial tip. The same tips were also used to image the AC-erased region of the hard drive (Fig. 5(a) and (b)). The internal structure of the bits was better resolved with the CCNT tip (Fig. 5(b)). The area around the tracks showed irregular features smaller than 20nm, which were presumably created by the AC-erase process.

A section line between points A and B on the 800kfci track (Fig. 5(c)) shows clear resolution of 18 nm features. A Fourier transform of the same section line taken across the entire image of Fig. 5(b) is shown in Fig. 5(d). It reveals a dominant spectral frequency in the bin centered at 1438 kfci, or 18 nm per flux change. These features are smaller than the 32 nm per flux change expected for an 800 kfci track. We do not understand the reason for this frequency doubling in the tracks written on the AC-erased area, which was not observed in tracks written on DC background. The frequency doubling has been observed in the AC-erased area of the sample with two separate tips. Each of these tips was nominally coated with a 3 nm Ti sticking layer, a 7 nm Co magnetic layer, and a 3 nm Ti cap layer, from the direction indicated in Fig. 1, and each had a diameter of less than 20 nm.

We have shown that metal-coated carbon nanotube tips can be conveniently produced with controlled specifications and that they can be used to resolve magnetic features smaller than 20nm. The use of nanotube tips should also help the development of MFM tips that can be operated on high aspect ratio features, and may help in the quantitative analysis of MFM data.

We thank Dennis Adderton of First Nano and Dr. Steve Minne of Veeco Instruments for the AFM probes, Prof. Shan X. Wang of Stanford University and Dr. Davide Guarisco of Maxtor Corporation for the recorded disks. This work was funded by AFOSR Grant F49620-00-1-0315, by the NSF under grant DMR-0103548, DARPA-MTO Lithography Network and by the Packard Foundation.

**Captions for figures**

FIGURE 1. Scanning Electron Microscope (SEM) image of an MFM cantilever. Inset, upper left: high-resolution SEM image of the apex of the pyramid, where a coated carbon nanotube (CCNT) tip is visible. The arrow shows the direction

FIGURE 2: SEM images of 6 different tips with shortened, metal-coated carbon nanotube (CCNT) tips with varying coating thicknesses. Each image is on the same scale, with 100nm scale bars shown.

FIGURE 3: (a) Phase image of the 6 tracks with densities of 100kfci (254nm/fc), 300kfci (85nm/fc), 500kfci (51nm/fc), 700kfci (36nm/fc), 800kfci (32nm/fc), 1000kfci (25nm/fc) in the DC-erased area, taken with the CCNT tip in Fig. 2(a). The boundary between the AC and DC-erased areas appears at the top of the figure. (b) Zoomed in image of the 1000kfci track imaged with the same tip. In both images, the grayscale indicates phase shift in degrees. These images are comparable to those obtained by typical commercial tips.

FIGURE 4: (a) Phase image of the 700kfci, 800kfci and 1000kfci tracks in the DC-erased area, taken with a commercial tip (b) Phase image of the tracks with densities of 800kfci and 1000kfci in the DC-erased area, taken with the CCNT tip shown in Fig. 2(b). The grayscale indicates phase shift in degrees.

FIGURE 5: (a) Phase image of the 700kcfi and 800kfci tracks in the AC erased area, taken with the same commercial tip as in Fig. 4(a). (b) Phase image of the 700kfci and 800kfci tracks in the AC-erased area, taken with the same CCNT tip as in Fig. 4(b). The grayscale indicates phase shift in degrees. (c) A section line between the points shown by the arrows on the 800kfci track of (b). (d) A Fourier transform of the dotted line in (b).

Figure1 ; Zhifeng Deng ; Applied Physics Letters

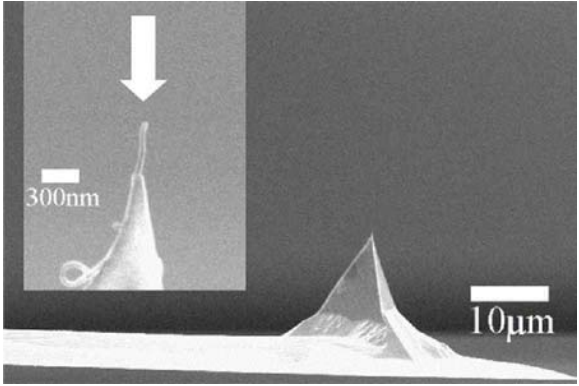

Figure2 ; Zhifeng Deng ; Applied Physics Letters

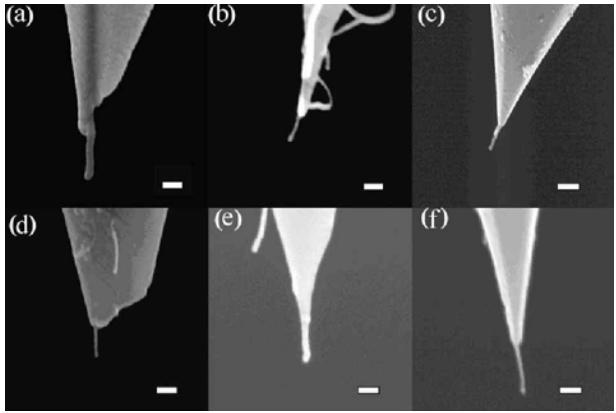

Figure3 ; Zhifeng Deng ; Applied Physics Letters

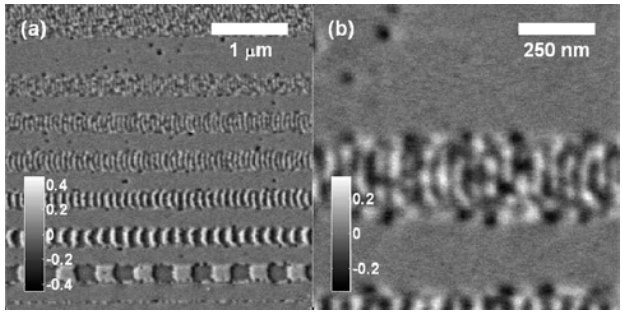

Figure4 ; Zhifeng Deng ; Applied Physics Letters

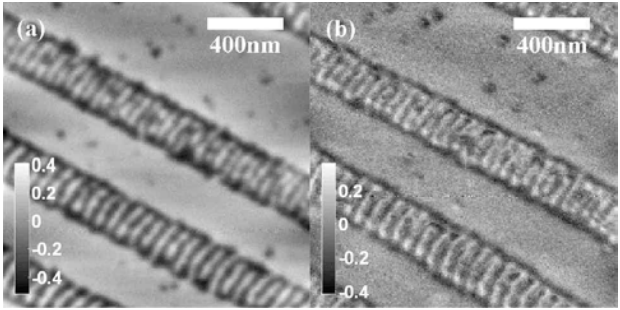

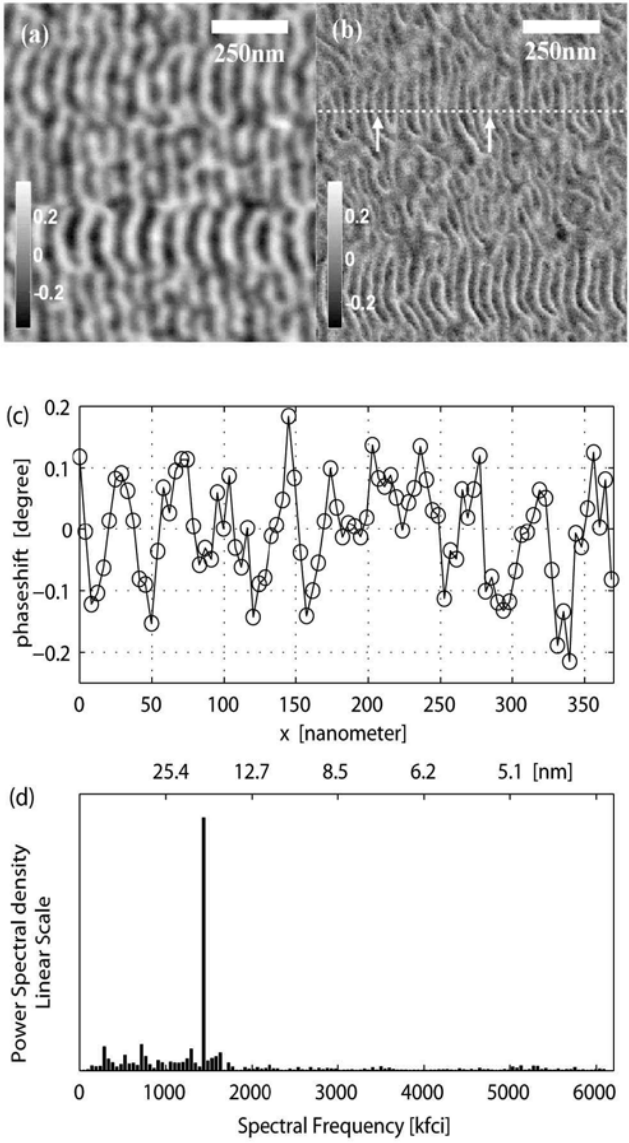